\documentclass{article}

\usepackage{graphicx}

\usepackage{amssymb}

\usepackage{amsmath}

\begin{document}

\title{A Nonlinear approach to measuring the effects of environmental variations}

\author{Mihaela D. Iftime \thanks{Boston, Massachusetts, May 1, 2010}}

\date{}

\maketitle

\begin{abstract}

This paper presents a nonlinear approach to measurements a 
general framework for dealing with variations of environmental conditions. 

My method may prove promising to extensions beyond classical physics, economics, and 
other sciences. I included few examples and applications of our method in Section 3 of this paper.

\end{abstract}

\pagebreak

\section{Statistical analysis of variations.}

As a result of the shift in attitudes toward the notion of equilibrium -- an old approach in sciences that treats systems 
as they are isolated, the study of environmental variability has become a subject of much interest.
Most research work are empirical studies related to problems in economic, social and health sciences concerns,
the results being disconnected because there is no theory to justify their use and results. 
Our theoretical results provide this missing part.

The paper uses the following terms defined as follows:
A reference system is referred to
an {\em open system}\footnote{ The notion of open system was originally formalized within the framework of thermodynamics.} that interacts with
its environment.
The {\em environment} consists of  all elements
outside a system, that has the potential to
affect the system.
The interaction with the environment may take many forms, such as: information, energy, etc, depending on the discipline it applies.
Changes in a system makes one aware of the presence of certain factors in the environment.
An {\em environmental factor}, represented by a
measurable variable $x$, that is a primary
environmental variable ( but it could be a higher level environmental variable derived from it).
Examples of environmental variables are:
heat (thermodynamics), fields (classical physics), the economy (business environment), etc.
I shall assume that there is a single {\em parameter of the system}, denoted $y$, that can changes
during various interactions with the environment.
For example, the temperature is a variable bound to the energy of a system that changes during heat transfer with the environment.

The section focuses on the construction of a strategy that can help one distinguish,
with a certain probability, real environmental effects.
I shall use the method of statistical inference to investigate a characteristic that is common to all environmental
influences, the change they bring about. The process can be divided into two main parts: 1) Data analysis and statement 
of the conjecture, and 2) Construction of a p-value statistics for testing the conjecture.\\

{\em Data preparation:}\\

The initial step is the examination of a large data representing a collection of values of 
the parameter $y$. If there are only small differences in the values of $y$, then this will often suffice to convince 
one that there is no need to further investigate the source of the observed variations. However, if upon examination, one
detects some wide divergences in the data, then further investigation is needed.
Step two is guess work: one should come up with a list of environmental conditions that can affect the data, outer factors
that must be measured or estimated.  Let assume $x$ is an environmental factor that
affects the parameter $y$ .\footnote{This paper only the one-dimensional case, however
our results are easily translatable to a multi-dimensional analysis of independent
(and dimensionally independent) environment variables.}
Suppose $\{x_{i}\}$, ${i=1,\ldots, n}$ are the
estimated values of $x$ that are related to
a set of values $\{y_{i}^{j}\}$, $j=1,\ldots, m_{i}$ (where
(${m_{i}\geq1}$ may differ for
different $i$). Another way is to interpret $x:Y\to\textbf{R}$ as a
real-valued function, where $Y$ denotes
the range space of variable $y$.

The collection $\mathcal{C}:= \{(x_{i},y^{j}_{i})\}$,
${i=1,\ldots, n}$, $j=1,\ldots, m_{i}$ of $\displaystyle N=\sum_{i, j}y^{i}_{j}$ paired values 
represents our empirical data to be analyzed.\\

{\em Sources of observed differences in the data:}\\

One can distinguish two sources for the differences in the $y^{i}_{j}$. First, there is a variation in $y$ that
is independent of the value of $x$. A measure of the
amount of inherent fluctuations in $y$
can be explained in terms of the entropy
$\displaystyle H(f)= -\sum_{i}{f_{i}log f_{i}}$ of the frequencies $f=\{f_{i}\}$ of $y^{j}_i$ distributed over different values $x_i$, for all ${i=1\ldots n}$.
The quantity, $H(f)\in [0,\mathrm{log}{n}]$; a value of $H(f)$ near $0$ indicates that $f$  is
associated to a highly predictable magnitude
of $x$.\footnote{This zero-entropic magnitude will define the left end-point of our scale interval constructed in Section 2.}
The second type of variations in $y^{i}_{j}$ are due to the differences in magnitudes of
the environmental factor $x$.  I want to determine statistically whether some observed
variation in the values of the parameter $y$ are within a range that could
signify an environmental effect, or it might have just happened by chance in the study.\\

{\em The conjecture: }\\

To do so, I shall make the following conjecture: ``The observed variation in the
values of the parameter $y$ at various values of
$x$ are real''.  To make judgments on the probability that the
observed differences are generated by the fluctuations
in the values of $x$, I shall test the conjecture using the method of inference.\\

{\em Testing the conjecture: }\\

In statistical inference, one indirectly tries to prove the conjecture by building up sufficient evidence to
disprove the contrary ( or the null
hypothesis $H_0$ statement).\footnote{The method of inference,
like the {\em reductio ad absurdum}, is an indirect proof of a conjecture to be false,
seeking to disprove the contrary statement (the null hypothesis $H_0$).}

In the language of frequencies, the probability of observing $f$ under the
 assumption that $H_0$ is true,
is given by the multinomial likelihood
$ L(f; p_{0}):= N!\displaystyle\frac{\prod_{i=1}^{n}p^{f_{i}}_{0i}}{\prod_{i=1}^{n} f_{i}!}$ of $f$ modulo a
"null-hypothetical" distribution $p_{0}$ (i.e., the distribution of the values of $y$
that would occur at each $x_i$ under the $H_0$ only).

The geometric average
$\displaystyle\phi^{2}=\sqrt[N]{L(f; p_{0})}$ is a
positive unknown quantity. $\phi^2\in[0,1]$ is independent of the number of measurements and represents the degree of deviation from
the null hypothesis. A value of $\phi^2$ near $1$ indicates that $f\rightarrow p_{0}$, whereas a value of $\phi^2$
near $0$ indicates that $f$ diverges from $p_{0}$. To be able to make statistical inferences about
our conjecture, one must estimate $\phi^2$ from the data. To do so, one must find an estimator of the null-hypothetical
distribution $p_0$, that is a
distribution of the values of $y$
that would occur at each $x_i$ with an
equal frequency. The closest uniform distribution that has
the expected mean\footnote{Note: When $x$ is a higher order environmental variables, higher central moments should be used. }
equal to the empirical average (i.e.,
 $\displaystyle\sum_{i=1}^{n}{x_{i}p_{0i}}=\sum_{i=1}^{n}x_{i}f_{i}$)
 is the maximum-entropy distribution estimator given by
 $\hat{p}_{0}(x_{i})=\hat{p}_{0i}:= ce^{\lambda x_{i}}$ where $i=1\ldots n$ where $c=$ const., and $\lambda=$
 Lagrange multiplier.\footnote{The existence and uniqueness of maximum-entropy estimator $\hat{p}_{0}$ follows immediately from a result
 of Boltzmann\cite{Boltzmann} using calculus of variations} The method of maximum entropy
was employed to determine a null-hypothetical estimator
 $\hat{p}_{0}$ that satisfy the property of
maximizing the Shannon (information-) entropy
$H(\hat{p}_{0})= H_{max}$ over the space of frequency
distributions with support in $Y$ and having the expected mean close to the ''empirical'' average.\footnote{The mean-value condition translates the fact that, under $H_0$ assumption one ignores
the association between $y$ and $x$}

By applying Stirling's approximation in the limit
 $N\to\infty$, an estimated value for the geometric average log-likelihood is given by:
$\mathrm{log}\phi^2(f;\hat{p}_{0})\approx -
 d_{KL}(p\|\hat{p}_{0})$ where
 $\displaystyle d_{KL}(p\|\hat{p}_{0})= \sum_{i}^{n}{p_{i}\mathrm{log}\frac{p_{i}}{\hat{p}_{0i}}} = \mathrm{log} n  - H_{max}$
 denotes the Kullback-Lieber measure of the
 divergence of $\hat{p}_{0}$ from $p$, the ''empirical'' probability, where
  $\displaystyle p_{i}=\mathrm{lim}_{N\to\infty}f_{i}$.
The estimated value of $\mathrm{log}\phi^2$ measures the degree of the difference
between the observed and the theoretical distributions of the values of $y$ at different values of $x$, and
it can be used to define a test statistics, $|TS|: =N\mathrm{log}\phi^2(f; \hat{p}_{0})$.\\

{\em P-value:}\\

To check whether the null-hypothesis is plausible, one can compare the value of the test
statistics $|TS|$ to a $\chi^2$ distribution with $(n-2)$
degrees of freedom \cite{Pearson}. A statistical test of our conjecture is as follows:

If the {\em p-value}$=2\mathrm{Prob}\{\chi^{2}_{n-2}\geq |TS|\}$ is smaller than or equal to
$\alpha$, the null-hypothesis must be rejected.  In other words, our conjecture is favored, if the absolute value of the test statistics, $|TS|$, is greater than the critical value $c_{n-2,\alpha}$
corresponding to a desired $\alpha$. \footnote{The "correct" significance level to be used in practice often depends on the case of study}
In this case, I found a strong statistical relationship
between the fluctuations in the environmental factor $x$ and the observed differences
in the values $y^{j}_i$. A 'real' proof of the conjecture often requires further investigation.

The cutoff value $c_{n-2,\alpha}$ can be used to provide a
prediction interval: Approximately a
proportion $(1-\alpha)$ of the total observed differences in $y^{i}_{j}$ that can be explained by differences in
the magnitudes of $x$, will have
$|TS| =N\mathrm{log}\phi^2(f; \hat{p}_{0})$ within the prediction interval $\displaystyle[0,c_{n-2,\alpha}]$.

\section{Nonlinear numerical representation of environmental effects.}

The majority of quantitative measurement treatments in the physical sciences and engineering use a linear approach, that
begins with a particular choice of frame of coordinates -- a
scale representation, with units as standards for measurements.
This section presents a nonlinear method for quantifying the environmental effects.

For the purpose of constructing our mathematical model,
I shall assume the following desirable characteristics of
the environmental factor: There is {\em latency period}, i.e.,
a lapse of a certain time interval taken by
an environmental factor to produce a change in a system.
There is an {\em lowest} $x_1$ and an
{\em upper} $x_m$ limit magnitude, below and above
which the environmental factor $x$ induces almost indefinite changes.
There is an approximate {\em proportionality
relationship} between an increase change
in the magnitude of $x$ and the related increase change in the values of $y$.

I shall construct a nonlinear approach to measuring the observed changes in the parameter $y$ caused by fluctuations
in the environmental factor that affects the system.
The measurement process consists of two main steps: 1)Construction of a geometric representation for
relative changes, and 2) Nonlinear numeric scale.\\

{\em Analytic non-linear geometric representation:}\\

There are two common usages for numbers:
counting and measuring.
Our mathematical intuition tells us
that the two thinking abilities
require quite different approaches.
In measuring, one uses numbers in order
to discriminate between changes
in qualitative or empirical attributes of
variables that are measured.
Our approach to measuring the observed changes in $y$ 
caused by the fluctuations in the environmental factor, is based on the construction of a nonlinear analytical geometric
model.

Let us go back to our data set $\{(x_{i},y^{j}_{i})\}$, ${i=1,\ldots, n}$
and compute the average values $\bar{y}_i$ of $\{y_{i}^{1},\ldots,y_{i}^{m_{i}} \}$ for
at each $i=1\ldots n$. The new collection  $\displaystyle\mathcal{C'}:= \{(x_{i}, \bar{y}_{i})\}$,
${i=1\ldots n}$ will be further used in our analysis. One can also interpret
$\bar{y}(x): [x_{1},x_{m}]\to \mathbf{R}$
 as a function, $\bar{y}(x_{i})=\bar{y_{i}}$ define for all $x_i$ inside the threshold interval $[x_{1},x_{m}]$.

Relative changes in $y$ generated by fluctuations 
$d_{ij}x=|x_{i}-x_{j}|$, $i\neq j =1,\ldots, n$ in the environmental factor $x$ are defined in
terms of the relative differences between the averages,
$\delta_{ij}y=\frac{\bar{y}_{i}-\bar{y}_{j}}{\bar{y}_{i}}$. It is clear that for any two values $x_{i},x_{j}\in[x_{1},x_{m}]$ there is a unique relative change $\delta_{ij}y$.

One can easily visualize relative displacements $\delta y$ as ``free''
vectors $w$ of magnitude $|w|=\delta y$. To be able to find a model for the space of free vectors, one must notice the following properties:
the space of vectors has a distinguishable structure given by the relative distance, it allows for vector
addition (i.e., a succession of relative displacements is achieved by an addition of vectors) and that there is an
vector $w_{0}\neq 0$ of magnitude $|w_{0}| =\mathrm{min}_{i,j}\{\delta_{ij}y\}$, the smallest relative change
generated by the measurable environmental factor fluctuations.

Since in the mathematics context one can only talk about vectors as ``directed line segments'' originating at the
same origin, I shall bound the ``free" vectors at the same origin that is defined by the lowest magnitude $x_1$.
The resulting mathematical formalization of the
space of vectors satisfying the properties: it has a commutative addition operation, an initial point, and a distinguishable
structure given by the "relative distance", is the abstract cyclic group $\mathbf{Z}_{m}= <r>$ generated by 
$r=\mathrm{min}_{i,j}\{\delta_{ij}y\}$ -- the smallest relative
change in the system generated by the environmental factor fluctuations.\footnote{Here we implicitly assumed that the relationship between x and y variables is approximately stable,
such that $r$ remains relatively constant over the interval $[x_{1},x_{m}]$.}.
In this abstract mathematical model, relative changes $\delta y$ in $y$ generated by fluctuations $dx$ of the
environmental factor $x$ have the geometric representation as vectors defined abstractly as points of $\mathbf{Z}_{m}$.
When the cyclic group is written multiplicatively, every element $z_{k}\in\mathbf{Z}_{m}$ can be 
written as power of $r$. So for any relative displacement $|w|=\delta y$,
 there is a $1\leq k\leq m$ such that $\delta{y}=r^k$. For any two vectors $|w_{1}|=r^{k_1}$,
 $|w_{2}|=r^{k_2}$, the sum is given by $|w_{1} + w_2|=r^{k_{1}+ k_{2}}$.
 
Using this geometric representation, I shall go on now to construct a one-dimensional numeric scale for measuring
the effect of environmental factor fluctuations.\\

{\em Nonlinear numeric scale: }\\

The scale construction is as follows. The elements $z_{k}\in\mathrm{Z}_{m}$ are arranged on a
positive axis, that defines the scale interval $[0, z_{m-1}]$, where the left end-point $0$ 
is given by the zero-entropic magnitude of the environmental factor, while the right end-point is
determined by the order $m$ of the cyclic group model. The relative changes are represented as 
points on the scale interval; each $\delta y$ has a unique scale representation, as
a power of $r$. It means that $r$ is the ``relative unit'' for measurements on the nonlinear scale.
One can also interpret $r$ as the nonlinear geometric generalization of the notion of one-dimensional linear basis( a ``free" mathematical 
object used as the metric unit in linear measurements). A relative unit reveals the structure of the
cyclic group model in a concise way, but it is not uniquely defined. A notion of ``relativity'' can be
defined as follows: two systems (belonging to the same class) with different relative units $r\neq r'$ will 
experience similar effects. A relative change in the systems' parameters $y$ and $y'$ generated by a
fluctuations in the environmental factor are related by a scaling similarity factor $s=\frac{r}{r'}$. The cyclic group
 $\mathrm{Z}_m$ is the same (isomorphic). The order $m$ of the cyclic group
defines the range of measurable magnitudes; the analog of dimension for linear spaces.
A cyclic group is the standard representation for measurements of changes in systems produced by fluctuations in the
environment that affects all ``similar'' systems in a similar way.  
Our nonlinear model is not data dependent, though it emerged from the analysis of empirical properties of the interaction of the system with its
environment.

On a nonlinear scale it is easy to make comparison between different changes produced by an environmental variable, where 
numbers are used to represent proportions. For example, two different changes $\delta_{1}y$, $\delta_{2}y$ having nonlinear 
scale expressions $\delta_{1}y=r^{k_1}$, $\delta_{2}y=r^{k_2}$. If easy to see that e.g., $\delta_{1}y\geq\delta_{2}y$ if
 $k_{1}\geq k_2$, etc. 

The nonlinear scale is a useful tool for making predictions of future changes in the system parameter at
various environmental conditions. For example, if one wants to measure the smallest change in the values of a 
parameter $y$ that is greater than $5$ on a given nonlinear measurement scale $[0, 100]$ with relative unit $r=2$.
It is easy to prove that the predictable value is $15$. Indeed, one
can easily check this by using the computation for $r = 2 =(15-5)/5$.
Secondly, to compare two changes $(5, 15)$ and $(15,30)$ is to appreciate
the proportional change between them. It is easy to appreciate with precision
that the produced environmental effect is the same, since $(30-15)/15=2=(15-5)/5 $.\\

\section{Examples and Applications.}

\subsection{Temperature measurements.}

An important example in the physical sciences is temperature measurements. Temperature is one of the main parameters
of a physical system that is used to describe the thermal mechanism transforming heat from the surrounding to the system.
Using thermodynamic methods, one can construct a nonlinear scale for temperature measurements, known as 
the "universal" relative ratio temperature scale. From the second law of thermodynamics, one determines
that the smallest relative difference in temperature $\frac{T_{2}- T_{1}}{T_{1}}$ 
is the efficiency coefficient,  a universal constant independent of the nature of the working substance
or the type of energy that is the source of work. The nonlinear temperature scale has a zero-entropy origin,
that is the absolute zero temperature value.\cite{temp}

\subsection{Financial Forecasting.}

Most often used financial forecasting techniques are time series and regression.
Time series analysis takes into account "noise" and other trends, such as seasonality, but it doesn't deal with outer factors. 
Regression is a statistical analysis is considered the most accurate forecasting method available.\cite{Mentzer} 
An important limitation of the regression model is that it assumes only the analysis of absolute variations and that the relationship
between the variable is stable(linear), which often this is not the case. None of the existing methods consider environmental variables selection based on how it affects the variation in $y$ when $x$ is taken into account.

Our nonlinear method can be applied to analyze how environmental factors affect the fluctuations 
in customers demand ( sales). In that case, $y$ represents the demand (sale) variable, 
and $x$ represents an environmental factors. Examples of environmental factors are: the economy, government rules and regulations 
(change in these regulations can impact the decisions of the company), political stability 
(can have a huge impact on the operations of businesses), promotional expenditures, suppliers availability (can positively or
 negatively affect a company, e.g., the availability of material and natural resources can impact the core business of companies), and the 
overall environment and culture ( which can also have a huge impact on a business).

Our non-linear numeric scale for measurements of the fluctuations in demand (I call it, an {\em oicometer}) can be used as a financial 
instrument for the analysis (comparison) of percentages of changes in sales, as well as to forecast the sales trend under changing environmental 
conditions. 

The fluctuating economic reality seems to point to the idea that price equilibrium is rather impossible.
Our nonlinear approach provides an
explanation of how much customers are willing to pay for a desired product in the context of a 
competitive market.\cite{Iftime2010b}

\section{History and Implications}

Linear models have a long tradition in mathematics, and they have been used as a standardized
quantitative approach in all sciences.

Historically, the notion of linear space grew up from affine geometry,
via the introduction of coordinates in a plane. In the 17th century, Descartes and Fermat founded the
analytic geometry - first time when solutions to an equation of two variables  were identified with points on
a plane curve.\cite{Bourbaki} Today linearized methods (including infinite-dimensional cases) are applied throughout mathematics and the
physical science. Linear methods are becoming a firmly
established numerical measurement tool, 
and furnishing the tensorial language representation for gauge and gravitations physical fields,
as well as providing an environment for Lebesque's construction of function spaces -
the background for a variety of solution techniques for partial differential equations.

In a ''linearized world'', quantitative attributes are being measured on a real scale axis, in which a
difference between the levels of an attribute are multiplied by any real number to exceed or equal another difference.
Measurement are estimations of
absolute ratios between the magnitude of a continuous quantity and a unit magnitude of the same kind.
Mass, length, time, plane angle, energy and electric charge are such examples. The majority
of existing numeric approaches treat only weak
nonlinearities or are corrections to behaviors that are distortions of linear behavior.
The main reason to continue to apply linear methods is to make use of a broad variety of developed techniques; e.g.,
 all statistical measures
can be used for a variable measured at linear scale levels, as all necessary mathematical operations are defined,
linear numeric scale representation. A major concern is when one tries to recover the
underlying qualitative relationship from an unsuitable numeric model. For example, its application ``creates" chaotic evolutions: one can generate fractals by simply iterating a collection of affine transformations
(IFS method) of the plane (a composition of scaling, reflections, rotations, and translations,
in which the order of these transformations is important).
Since the evolution equations of fractals are a set of scale invariant rules, a nonlinear approach
seems more suitable for dealing with objects that evolve proportionally.

Our nonlinear approach is useful alongside the linear methods. As I have shown in this paper, 
the analysis of both absolute and relative changes are important, since either one alone might be misleading. 
It is easy to see that
a wide divergence in the absolute differences that result in only a small
relative change (and the other way around) may not be relevant for making predictions.

Our nonlinear numeric representation formalizes a natural way
of comparing numeric magnitudes. Extensive literature in the cognitive neuroscience and brain-imaging research
show that humans tend to perform better and faster on
comparison tasks when using proportions of visual
images, rather than the computing in terms of ordinary linear scale representation.
In fact, in Galton's experiment none of his subjects reported a linear number axis.\cite{Galton}

The two properties of our nonlinear model, associativity and similarity, have found an analog in the
physical characteristics of our long-term memory, associativity and content-addressability (the ability to
readily retrieving related content based on similarity), as many experimental studies suggest as related to
visual numeric representation.\cite{Dehaene}

It is interesting to point that a ``visual" geometrical representation of numbers can provide an explication to the
Plato's view that numbers have their own independent reality.


\begin{thebibliography}{100}

\bibitem{Boltzmann} L. Boltzmann,\textit{The second law of thermodynamics}, 1974

\bibitem{Bourbaki} Bourbaki,\textit{Algebre linéaire et algebre multilinéaire}, 1969

\bibitem{Brannon} K. E. Jordan, E.M. Brannon,\textit{A common representational system governed by
Weber's law: Nonverbal numerical similarity
judgments in 6-year-olds and rhesus macaques}, J.of Experimental Child Psychology 95, 2006

\bibitem{Dehaene} S. Dehaene, E. Spelke, P. Pinel, R. Stanescu, S. Tsivkin,\textit{Sources of Mathematical
Thinking: Behavioral and
Brain maging Evidence, Science} 284, 970, 1999

\bibitem{Galton} F. Galton, \textit{Questions on the faculty of visions}, 1897

\bibitem{Iftime2010b} M. D. Iftime, \textit{A nonlinear forecasting model of demand fluctuations}, submitted  J. of Forecasting, 2010

\bibitem{temp} S. I. Kolesnikov, V. A. Vinokurov, G. E. Zaikov, I. M. Kolesnikov,
\textit{Thermodynamics of spontaneous and non-spontaneous processes}, Nova Sci. Publishers, 2001
  
\bibitem{KL} S. Kullback, \textit{The Kullback-–Leibler distance}, The American Statistician 41,(4),1987

\bibitem{Longo} M.R. Longo, S.F. Lourenco ,\textit{
Spatial attention and the mental number line: Evidence for characteristic biases and compression.}
 Neuropsychologia, 45 (7), 2007

\bibitem{Mentzer} J. Mentzer, Sales forecasting management, in H. E. Kyburg and H. E. Smokler (eds), Studies in Subjective Probability, New York: Wiley, 1964.


\bibitem{Pearson} K. Pearson, \textit{On the criterion that a given system of deviations from the probable in
the case of a correlated system of variables is such that it can be reasonably supposed to have arisen from random
sampling.} Phil. Mag. 50, (5), 1900




\end{thebibliography}
\end{document}